\documentclass[pra,twocolumn,floatix,amssymb,showpacs,amsmath,superscriptaddress]{revtex4-1}

\usepackage{graphicx}
\usepackage{epsfig}
\usepackage{amsmath,amssymb}
\usepackage{ dsfont }
\usepackage{color}
\usepackage[utf8]{inputenc}
\usepackage[T1]{fontenc}
\usepackage[english]{babel} 
\usepackage{float}
\usepackage{times}
\usepackage{dsfont}
\usepackage{hyperref,amsmath,amssymb,amscd}
\usepackage{amssymb,amsmath,amsfonts,amsthm}
\usepackage{mathtools}
\usepackage{verbatim}
\usepackage{enumerate}
\usepackage{graphicx}
\graphicspath{{figs/}}
\usepackage{hyperref}
\usepackage{bbm}
\usepackage{booktabs}

\usepackage[caption=false]{subfig}

\usepackage{color}
\definecolor{dred}{rgb}{.8,0.2,.2}
\definecolor{ddred}{rgb}{.8,0.5,.5}
\definecolor{dblue}{rgb}{.2,0.2,.8}
\definecolor{dgreen}{rgb}{.2,0.5,.2}

\newcommand{\bra}[1]{\mbox{$\langle #1|$}}
\newcommand{\ket}[1]{\ensuremath{|#1\rangle}}

\newcommand{\be}{\begin{equation}}
\newcommand{\ee}{\end{equation}}

\newcommand{\bea}{\begin{eqnarray}}
\newcommand{\eea}{\end{eqnarray}}

\begin{document}

\newcommand{\quadr}[1]{\ensuremath{{\not}{#1}}}
\newcommand{\quadrd}[0]{\ensuremath{{\not}{\partial}}}
\newcommand{\slpar}{\partial\!\!\!/}
\newcommand{\gtrescero}{\gamma_{(3)}^0}
\newcommand{\gtresuno}{\gamma_{(3)}^1}
\newcommand{\gtresi}{\gamma_{(3)}^i}

\title{Experimental Simulation of Bosonic Creation and Annihilation Operators in a Quantum Processor }

\author{Xiangyu Kong}
\thanks{These two authors contribute equally to this study.}
\address{ State Key Laboratory of Low-Dimensional Quantum Physics and Department of Physics, Tsinghua University, Beijing 100084, China}

\author{Shijie Wei}
\thanks{These two authors contribute equally to this study.}
\address{ IBM Research, China}

\author{Jingwei Wen}
\address{ State Key Laboratory of Low-Dimensional Quantum Physics and Department of Physics, Tsinghua University, Beijing 100084, China}

\author{Gui-Lu Long}
\email{Correspondence and requests for materials should be addressed to G.L.L.:gllong@tsinghua.edu.cn}

\address{ State Key Laboratory of Low-Dimensional Quantum Physics and Department of Physics, Tsinghua University, Beijing 100084, China}

\address{ Tsinghua National Laboratory for Information Science and
Technology, Beijing 100084, P. R. China.}

\address{ Collaborative Innovation Center of Quantum Matter, Beijing 100084, China}

\date{\today}

\begin{abstract}
The ability of implementing quantum operations plays fundamental role in manipulating quantum systems. Creation and annihilation operators which transform a quantum state to another by adding or subtracting a particle are crucial of constructing quantum description of many body quantum theory and quantum field theory. Here we present a quantum algorithm to perform them by the linear combination of unitary operations associated with a two-qubit ancillary system. Our method can realize creation and annihilation operators simultaneously in the subspace of the whole system. A prototypical experiment was performed with a 4-qubit Nuclear Magnetic Resonance processor, demonstrating the algorithm via full state tomography. The creation and annihilation operators are realized with a fidelity all above 96\% and a probability about 50\%. Moreover, our method can be employed to quantum random walk in an arbitrary initial state. With the prosperous development of quantum computing, our work provides a quantum control technology to implement non-unitary evolution in near-term quantum computer.
\end{abstract}


\maketitle

\section{INTRODUCTION}
The non-Hermitian bosonic operators $\hat{a}$ and $\hat{a}^\dagger$ introduced in harmonic oscillator question is a basic and crucial concept in quantum mechanics, laying the foundation for
quadratic quantization\cite{sakura}. They offer us an alternative way to calculate harmonic oscillator system without solving the irksome differential equations\cite{diff1,diff2}. These bosonic operators also play an important role in many fields of physics such as quantum optics\cite{optical}, quantum mechanics\cite{sakura,mechanic}, quantum measurement\cite{measure} and even
quantum chemistry\cite{chemistry}. According to existing research, the annihilation and creation operation can be foundations to construct arbitrary quantum state in theory\cite{state}. Considering the
hot field of quantum computation and quantum information,
it is natural to for us try to realize these operators in
steerable quantum system which can provide us a novel
way to design quantum algorithm. The designed quantum system is difficult to execute non-unitary operator, which means there will be an obstacle to evolve the max and min quantum state to zero state in the process of realizing creation and annihilation operators.
Given the importance of these operators, efficiently performing them with high success probability and
fidelity in quantum process is critical. Recently, we realize some experiment progress and some works focus on realizing the bosonic operations at the single-boson level in optical
system\cite{optical}. But the probability of success and performance
fidelity are things belong to coin’s different sides in general.
Some improvements have been done in trapped ion system realizing deterministic addition and near-deterministic subtraction
of bosonic particle with fidelity over 0.9\cite{jin}. However, improvements
still need to be made to satisfy higher precision and less experiment times demanded.

In this paper, we experimentally realize creation and annihilation operators using the combination of unitary operators in a four-qubit nuclear magnetic resonance (NMR) system for the first time. The results offer
us a raise both in experiment precision and success ratio. The paper is organized as follows: In Sec. II, we introduce
the universal theory of how to realize these two non-unitary
operators. In Sec. III, we take the four-qubit sample as an example to introduce our experimental setups and experimental procedure.
Then, we present the experimental results and discuss the consequences. In Sec. IV, we report an application of our algorithm. At last, we close with a conclusion section summarizing the entire work and giving some prospects.

\section{Theory}

Quantum mechanically, creation $\hat{a}^\dagger$ and annihilation $\hat{a}$ operators acting on a bosonic system with number $N$ of identical
particles satisfy the following operator relationship\cite{dagger}:
\begin{eqnarray}
\hat a^\dag\ket N &=& \sqrt{N+1}\ket{N+1} \\ \nonumber
\hat a\ket N &=& \sqrt{N}\ket{N-1}.
\end{eqnarray}
Given the importance of the creation and annihilation operators, efficiently performing them in a quantum process is critical.
However, implementing such bosonic operations is challenging. Because the fact that these operators are non-unitary and inherently probabilistic, cannot be realized during the Hamiltonian evolution of a physical system without enlarging the Hilbert space. 

Ignoring the modification of the probability amplitude of state, the conventional addition and subtraction of a particle can be expressed as
\begin{eqnarray}
\hat K^\dag\ket N &=& \ket{N+1} \\ \nonumber
\hat K\ket N &=& \ket{N-1}.
\end{eqnarray}
which can be called addition and subtraction operation.

We consider addition and subtraction operators with identical particles number $ N $. Our method can realize addition $ \hat K^\dag$ and subtraction $ \hat K $
operators in one quantum circuit.

In our method, the identical particles number $ N $ is mapped to a corresponding $\ket{N } $ state. For convenience, we adopt a truncated form of creation operation $ \hat K^\dag$ by defining it as a $ (N+1) \otimes (N+1) $ matrix
\begin{eqnarray}
\hat{K}^{\dagger}=\sum\limits_{i=0}^{N-1}\ket{i+1}\bra{i},\quad
\hat{K}=\sum\limits_{i=1}^{N}\ket{i-1}\bra{i} \label{3}
\end{eqnarray}
Every operation above can be decomposed into a sum form of two unitary operations.
\begin{eqnarray}
\hat{K}^{\dagger}=\frac{U_{0}+U_{1}}{2},\quad \hat{K}= \frac{U_{2}+U_{3}}{2}, \label{4}
\end{eqnarray}
where
\begin{eqnarray}
U_{0}&=&\sum\limits_{i=0}^{N-1}\ket{i+1}\bra{i}+ \ket{0}\bra{N}\nonumber\\
U_{1}&=&\sum\limits_{i=0}^{N-1}\ket{i+1}\bra{i}- \ket{0}\bra{N}\nonumber\\
U_{2}&=&\sum\limits_{i=1}^{N}\ket{i-1}\bra{i}+ \ket{N}\bra{0}\nonumber\\
U_{3}&=&\sum\limits_{i=1}^{N}\ket{i-1}\bra{i}- \ket{N}\bra{0}.\nonumber
\end{eqnarray}\label{5}

\begin{figure}
\centering
\includegraphics[width=0.45\textwidth]{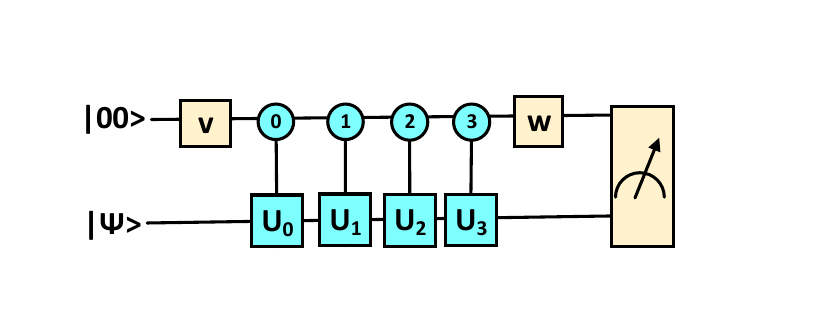}
\caption {Quantum circuit of realizing addition and subtraction operators. $|\Psi\rangle$ denotes the initial state of work system, and $|00\rangle$ is the initial state of auxiliary system. The squares represent unitary operations and the circles represent the state of the controlling qubit. Unitary operations $U_{0}$, $U_{1}$ , $U_{2}$ and $U_{3}$ are activated only when the auxiliary qubit is $|00\rangle$ , $|01\rangle$,$|10\rangle$ and $|11\rangle$ respectively. We "readout" two outputs with the first auxiliary system in state $ |00\rangle $, $ |10\rangle $. } \label{faa}
\end{figure}

Considering the fact $ \hat K^\dag$ and $ \hat K $ can be expressed into a linear combination of unitary operators, we can perform the addition and subtraction operators via duality quantum computing\cite{dqc1,dqc2,dqc3,dqc4}. In duality quantum computing, the work system with initial state $|\Psi\rangle$ and the $d$-dimension ancillary system with initial state $|0\rangle$ are coupled together. The corresponding digital simulation quantum circuit of the addition and subtraction operators by the algorithm is further shown in Fig. \ref{faa}.

As shown in Fig. \ref{faa}, the unitary operators $V$ and $W$ performed on the two-qubit ancillary system are
\begin{eqnarray}
V=H\otimes H\\
W=I\otimes H,
\end{eqnarray}
The circuit realizes the process:
\begin{eqnarray}
\label{process}
|0\rangle_{2}|\Psi\rangle & \rightarrow & \sqrt{\frac{1}{2}}\hat{K}^{\dagger}|00\rangle_{2}|\Psi\rangle + \sqrt{\frac{1}{2}}\hat{K}|10\rangle_{2}|\Psi\rangle \nonumber \\
& + & \sqrt{\frac{1}{2}}\hat{J}^{\dagger}|01\rangle_{2}|\Psi\rangle + \sqrt{\frac{1}{2}}\hat{J}|11\rangle_{2}|\Psi\rangle.
\end{eqnarray}
where the operators $\hat{K}^\dagger$, $\hat{K}$, $\hat{J}^\dagger$, $\hat{J}$ are acting on the work system $\ket{\Psi}$. To be the same with $\hat{K}^\dagger$ and $\hat{K}$ in the Eq. \ref{process}, the operators $\hat{J}^\dagger$ and $\hat{J}$ can be calculated by
\begin{equation}
\hat{J}^{\dagger}=\frac{U_0-U_1}{2}=\ket{0}\bra{N} , \quad \hat{J}=\frac{U_2-U_3}{2}=\ket{N}\bra{0}.
\end{equation}\label{3}
The $\hat{J}^{\dagger}$ operation transform $\ket{N } $ state into $\ket{0} $ state, and $\hat{J}$ transform $\ket{0} $ state into $\ket{N } $ state. So it is clear that the operator $\hat{J}^{\dagger}$ is a special addition operator dealing with the maximum state and the operator $\hat{J}$ is the special subtraction operator dealing with the minimum state. The probability of realization of addition operator and subtraction are both 50\%.


Then we will consider addition and subtraction operations in the case where identical particles number $ N=3 $ in Eq. \ref{4}.

\begin{figure*}[htb]
\begin{center}
\includegraphics[width=2.1\columnwidth]{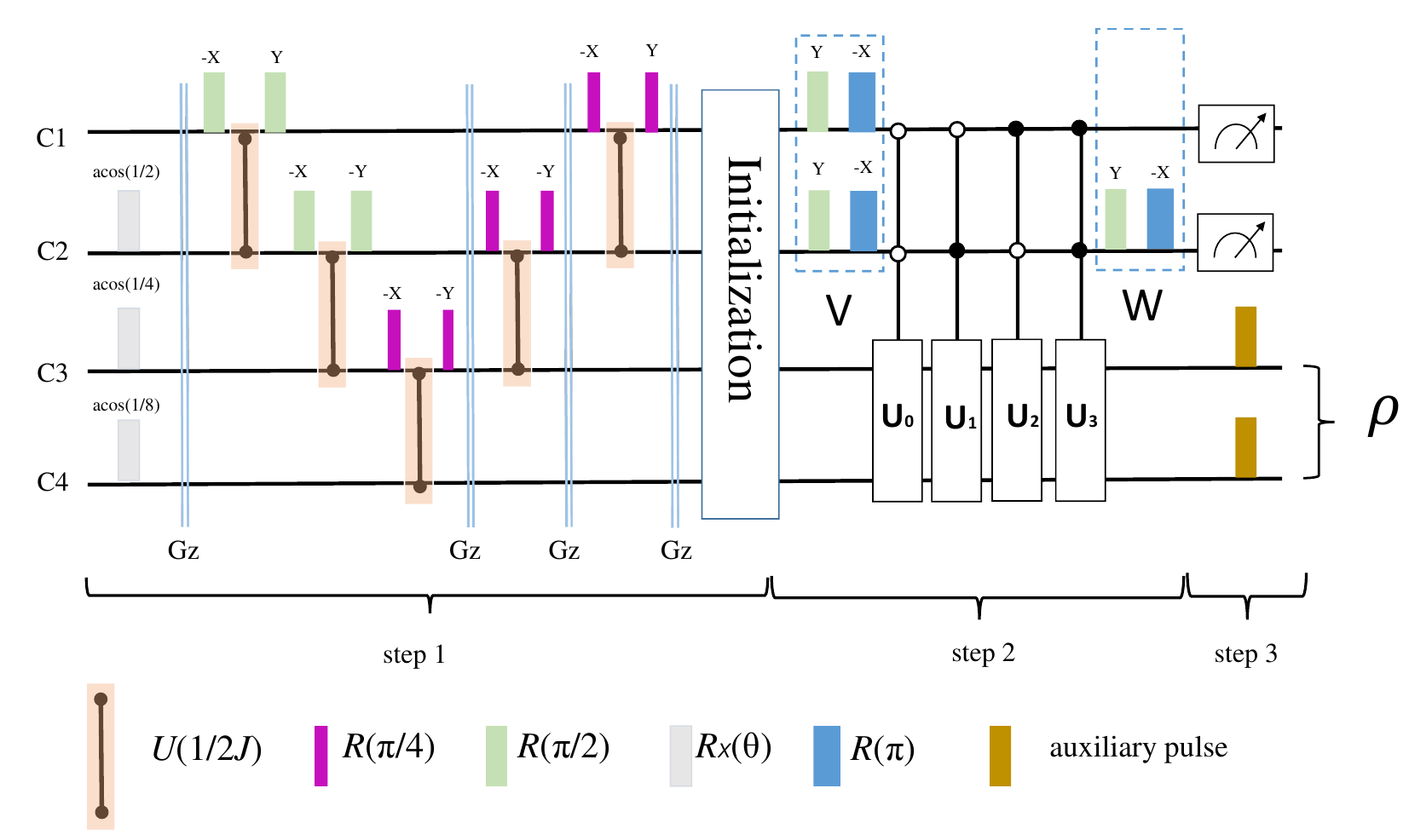}
\end{center}
\setlength{\abovecaptionskip}{-0.00cm}
\caption {\footnotesize{\textbf{NMR sequence to realize the addition and subtraction operators.} In the first step, we use spatial averaging method to prepare the pseudo-pure state. In this part, the purple rectangles stand for the unitary operations which rotate forty-five degrees with a coordinate axis. Similarly, the green rectangles rotate ninety degrees. The white rectangles stand for the unitary operations which rotate the special angle with X axis. The blue rectangles stand for the unitary
operations which rotate one hundred and eighty degrees with Y axis. The double straight lines stand for the gradient field in the z direction. In the second step, the radio-frequency (RF) pulses during this procedure are optimized by the GRAPE technology to realize the controlled operator. In the third step, we measure the first two qubits and use the quantum tomography technology to obtain the final density matrix. The brown squares are the auxiliary pulse which can be used for quantum tomography.}}\label{fig:quantum circuit}
\end{figure*}

\section{experiment and result}
\label{experiment and result}
We experimentally inspect our algorithm using four-qubit nuclear magnetic resonance (NMR) syetem. As introduce before, the ancillary system is two qubits and we choose a two-qubit work system $\ket{00}, \ket{01}, \ket{10}, \ket{11}$ to represent 0,1,2,3.
The four-qubit sample is $^{13}$C-labeled trans-crotonic acid dissolved in d6-acetone.
\begin{figure}[htb]
\begin{center}
\includegraphics[width= 1\columnwidth]{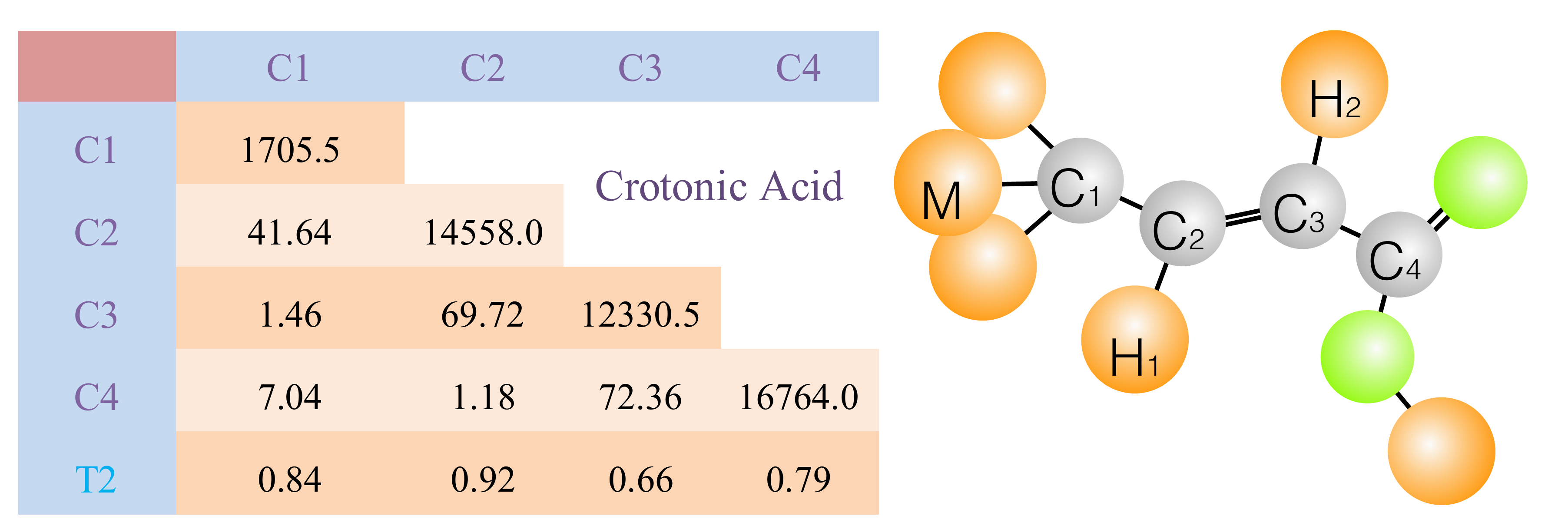}
\end{center}
\setlength{\abovecaptionskip}{-0.00cm}
\caption{\footnotesize{ Molecular structure and Hamiltonian parameters of $^{13}$C-labeled trans-crotonic acid. C$_1$, C$_2$, C$_3$ and C$_4$ are used as four qubits. The chemical shifts and J-couplings (in Hz) are listed by the diagonal and off-diagonal elements, respectively. T$_{2}$ (in Seconds) are also shown at bottom.}}\label{fig:molecule}
\end{figure}
The structure of the molecule is shown in Fig. \ref{fig:molecule}, where C$_1$ to C$_4$ denote the four qubits, the first two qubits are auxiliary system and the last two qubits are the work system. The methyl group M, H$_1$ and H$_2$ were decoupled throughout all experiments. For molecules
in liquid solution, both intramolecular dipolar couplings (between spins in the
same molecule) and intermolecular dipolar couplings (between spins in different molecules) are averaged away by the rapid tumbling\cite{nmr}. The internal Hamiltonian under weak coupling approximation is
\begin{align}\label{Hamiltonian}
\mathcal{H}=-\sum\limits_{j=1}^4 {\frac{1}{2} \omega _j } \sigma_z^j + \sum\limits_{j < k}^4 {\frac{\pi}{2}} J_{jk} \sigma_z^j \sigma_z^k,
\end{align}
where $\nu_j$ is the chemical shift and $\emph{J}_{jk}$ is the J-coupling strength. All experiments were carried out on a Bruker DRX 400MHz spectrometer at room temperature (296.5K).

The entire experiment can be divided into three parts and its experimental circuits are shown in Fig. \ref{fig:quantum circuit}.

\begin{figure}[htb]
\begin{center}
\includegraphics[width=1\columnwidth]{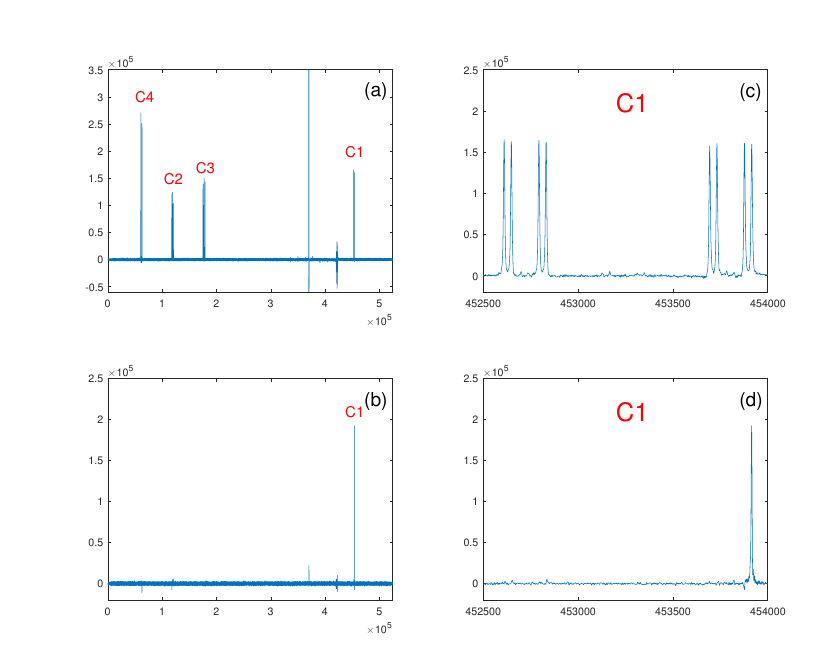}
\end{center}
\setlength{\abovecaptionskip}{-0.00cm}
\caption{\footnotesize{\textbf{The spectrum of thermal equilibrium state and PPS.} (a) The spectrum
of the thermal equilibrium state. The $x$ axis is the relative frequency and
the $y$ axis is the amplitude in arbitrary unit. Since the qubits are all carbon
atoms, all signal of these four qubits can be detected through carbon channel
in NMR experiments. (b) The spectrum of spin C1 in the thermal equilibrium
state. This is an expanded view of spin C1 in (a). From this figure, we can
observe more information in detail. Quantum state of remaining spins (C2,
C3 and C4) is as indicated, based on J14, J24 and J34 shown in Fig. 3. (c)
The spectrum of the PPS. As we know, the ground state \ket{0000} can not be
detected directly in NMR and we can apply the $R^i_y
(\pi/2)$ on it to obtain a
single peak of the $i$th spin. In our case, only the peak in the spectrum of C1
can be detected by applying $R^1_y
(\pi/2)$ pulse. (d) The spectrum of spin C1 in
PPS. This is also an expanded view of spin C1 in (c). It is clear that the peaks
standing for the quantum state except $\ket{0}$ are close to zero. }} \label{fig:pps}
\end{figure}

\textit{Step1:Initialization}---Starting from thermal equilibrium state, we drive the system to the pseudo-pure state (PPS) with the method of the spatial averaging technique~\cite{cory,D. Lu,Tao X,keren}. Step $1$ in Fig. \ref{fig:quantum circuit} is the experimental circuit realizing PPS where all local operations are optimized using the gradient ascent pulse engineering(GRAPE) with a fidelity over 99.5\%\cite{GRAPE1,GRAPE2}. The final form of four-qubit PPS is
$\rho_{0000}=(1-\epsilon){\mathbb{I}}/16+\epsilon\ket{0000}\bra{0000}$,
where $\mathbb{I}$ is identity matrix and $\epsilon\approx 10^{-5}$ is the polarization. Since only the deviated part $\ket{0000}$ contributes to the NMR signals, the density matrix used in NMR are all deviated matrix and the PPS is able to serve as an initial state. The experimental results are represented as the density matrices obtained by the state tomography technique~\cite{tomo1,tomo2,tomo3} shown in Fig. \ref{fig:pps}. The fidelity between the experimental results and $\ket{0000}$ is over 99.02\%.

\textit{Step2:operator}---As we introduced in the Sec. II, we perform the addition and subtraction operator $\hat{K}^{\dagger}$, $\hat{K}$ by the
duality quantum computing. In this case, C1 and C2 represent
the ancillary system with initial state $\ket{0}$, C3 and C4 represent the work system as shown in Fig.2. The unitary operator
V and W performed on the C1 and C2 can be realized as
shown in step2 at Fig.2. Moreover, the four Control-U operators can be realized by using the GRAPE technology with a fidelity over 99.5\%.

\textit{Step3:Mesurement}---
The measurement circuit is also listed
in Fig. 2 at Step 3. From Eq. \ref{process}, we know that the operator
applied on work system depends on the measurement of the
ancillary system. So we have to measure the ancillary qubits
and use quantum tomography technology to get the density
matrix by applying auxiliary pulses on C3, C4.

\section{result}
Here we will introduce two examples in detail with two different initial quantum states to show our algorithm. One is
general quantum state such as $\ket{\phi}=\ket{01}$, the other is superposition state such as $\ket{\phi}=\frac{\ket{01}+\ket{10}}{2}$. These initial states can
be obtained by single-qubit gates and C-NOT gate from the
ground state $\ket{00}$. Then we can get the experimental results shown below by following the steps introduced above. 

\subsection{general state}
\begin{figure}[htb]
\begin{center}
\includegraphics[width= 0.95\columnwidth]{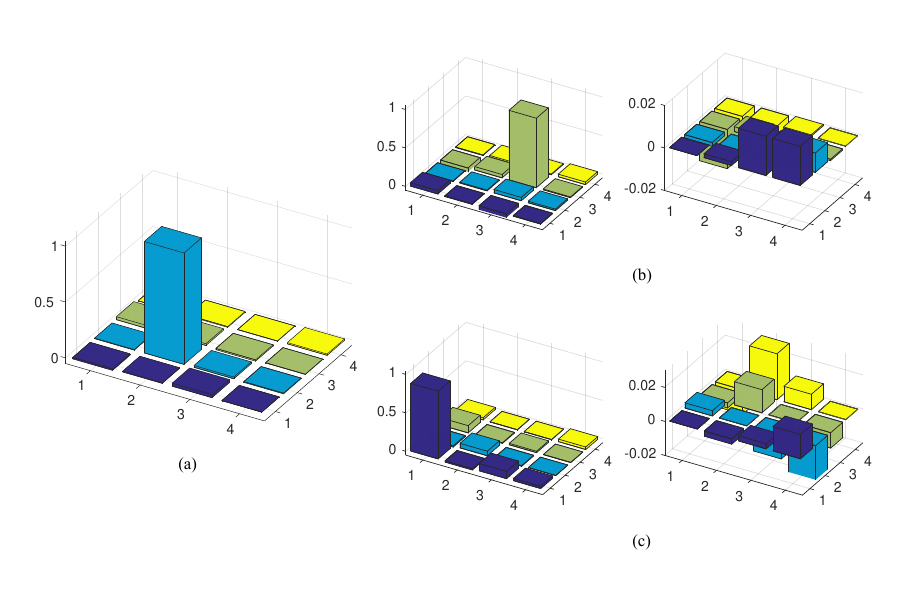}
\end{center}
\setlength{\abovecaptionskip}{-0.00cm}
\caption{\footnotesize{The density matrix of the initial quantum state and final state. (a) The initial quantum state $ \ket{01}$. (b) The final state when the measurement
of ancillary qubits is $ \ket{00}$. (c) The final state when the measurement of
ancillary qubits is $\ket{10}$. Moreover, the right part is the real part
of the density matrix and the left part is the imaginary part in (b) and (c).}} \label{result1}
\end{figure}

The experimental results of the first example are shown in
Fig.6. We can find that the addition operation can be realized
when the auxiliary qubits are $\ket{00}$. Then we calculate the fidelity
by definition\cite{fidelity}:
\begin{equation}
\label{fidelity}
F(\rho,\sigma)=|Tr(\rho \sigma)|/\sqrt{Tr(\rho^2)Tr(\sigma^2)}
\end{equation}
where $\rho$ and $\sigma$ represents the density matrix getting from the
experiments and theories respectively. So comparing the experiments
shown in Fig. \ref{result1}(b) with the quantum state $\ket{10}$, the
fidelity between them is over 98.8\%. Thus we probabilistically
realize the quantum process:
\begin{equation}
\label{add1}
\hat{K}^{\dagger}\ket{01}=\ket{10}
\end{equation}
Similarly, the subtraction operator can be performed when the
auxiliary qubit is $\ket{10}$ shown in Eq. 8. Comparing the results
shown Fig. \ref{result1} (c) with theoretical state $\ket{00}$, the fidelity
between them is approximately equal to 98.3\%. Thus we also
probabilistically realize the quantum process:
\begin{equation}
\label{sub1}
\hat{K}\ket{01}=\ket{00}
\end{equation}
We introduce that the addition and subtraction operators
can be performed in probability. From the Eq. 8, the probabilities
depends on the experimental results of the auxiliary qubit shown in Tab. II. The measurement of $\ket{11}$ should be
an experimental error. From the results we can find that we
can perform the addition and subtraction operator for the same
probability 50\%.

\begin{table}[tbp!]
\centering
\caption{\footnotesize{The measurement of the auxiliary qubit.}} \label{table:genneral}
\begin{tabular}{llll}
\hline
\hline
&\quad \quad $\ket{00}$ \quad \quad \quad \quad & $\ket{10}$ \quad \quad \quad \quad $\ket{01}$ and $\ket{11} $ \quad \quad \\
\hline
Parobability &\quad \quad 49.56\% \quad \quad \quad \quad & 49.51\% \quad \quad \quad \quad 0.93\% \quad \quad \\

\hline
\hline
\end{tabular}
\end{table}

\subsection{superposition state}
\begin{figure}[htb]
\begin{center}
\includegraphics[width= 0.95\columnwidth]{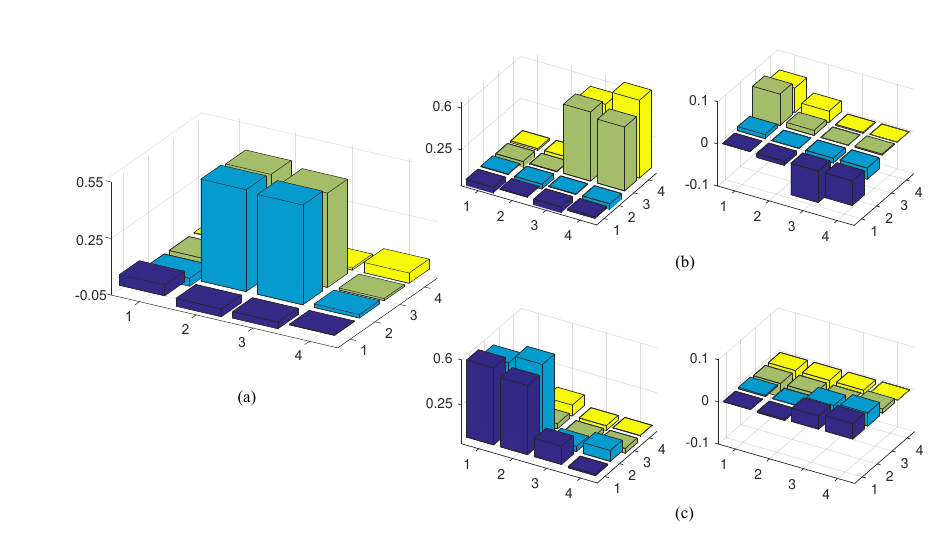}
\end{center}
\setlength{\abovecaptionskip}{-0.00cm}
\caption{\footnotesize{The density matrix of the initial quantum state and final state. (a)
is the initial quantum state $\frac{\ket{01}+\ket{10}}{2}$. (b) is the final state when the measurement
of ancillary qubits is $\ket{00}$. (c) is the final state when the measurement
of ancillary qubits is $\ket{10}$. Moreover, the right part is the real part of the density matrix and the left part is the imaginary part in (b) and (c).}} \label{result2}
\end{figure}

The experimental results of the second example are shown in
Fig. \ref{result2}. Similarly, we can find that the addition operation can be realized when the auxiliary qubit is $\ket{00}$. Then we calculate the fidelity by the definition Eq. 13.

So comparing the experiments shown in Fig. \ref{result2} (b) with the
quantum state $\frac{\ket{01}+\ket{10}}{2}$, the fidelity between them is over
96.3\%. Thus we realize the quantum process probabilistically:
\begin{equation}
\label{sub1}
K^+\frac{\ket{01}+\ket{10}}{2}=\frac{\ket{10}+\ket{11}}{2}
\end{equation}
Similarly, the subtraction operator can be performed when the
auxiliary qubit is $\ket{10}$ shown in Eq. 8. Comparing the results
shown Fig. \ref{result2}(c) with theoretical state $\frac{\ket{01}+\ket{10}}{2}$, the fidelity between
them is approximately equal to 97.0\%. Thus we also realize the quantum process probabilistically:
\begin{equation}
\label{sub1}
K^-\frac{\ket{01}+\ket{10}}{2}=\frac{\ket{00}+\ket{10}}{2}
\end{equation}
We introduce that the addition and subtraction operators
can be performed in probability. From the Eq. 8, the probabilities
depends on the experimental results of the auxiliary
qubit shown in Tab. II. The measurement of $\ket{11}$ should be an experimental error. From the results we can find that we
can perform the addition and subtraction operator for the same
probability 50\%.

\begin{table}[tbp!]
\centering
\caption{\footnotesize{The measurement of the auxiliary qubit.}} \label{table:superposition}
\begin{tabular}{llll}
\hline
\hline
&\quad \quad $\ket{00}$ \quad \quad \quad \quad & $\ket{10}$ \quad \quad \quad \quad $\ket{01}$ and $\ket{11} $ \quad \quad \\
\hline
Parobability &\quad \quad 48.84\% \quad \quad \quad \quad & 49.79\% \quad \quad \quad \quad 1.38\% \quad \quad \\

\hline
\hline
\end{tabular}
\end{table}

\section{application}
\begin{figure}[htb]
\begin{center}
\includegraphics[width= 0.95\columnwidth]{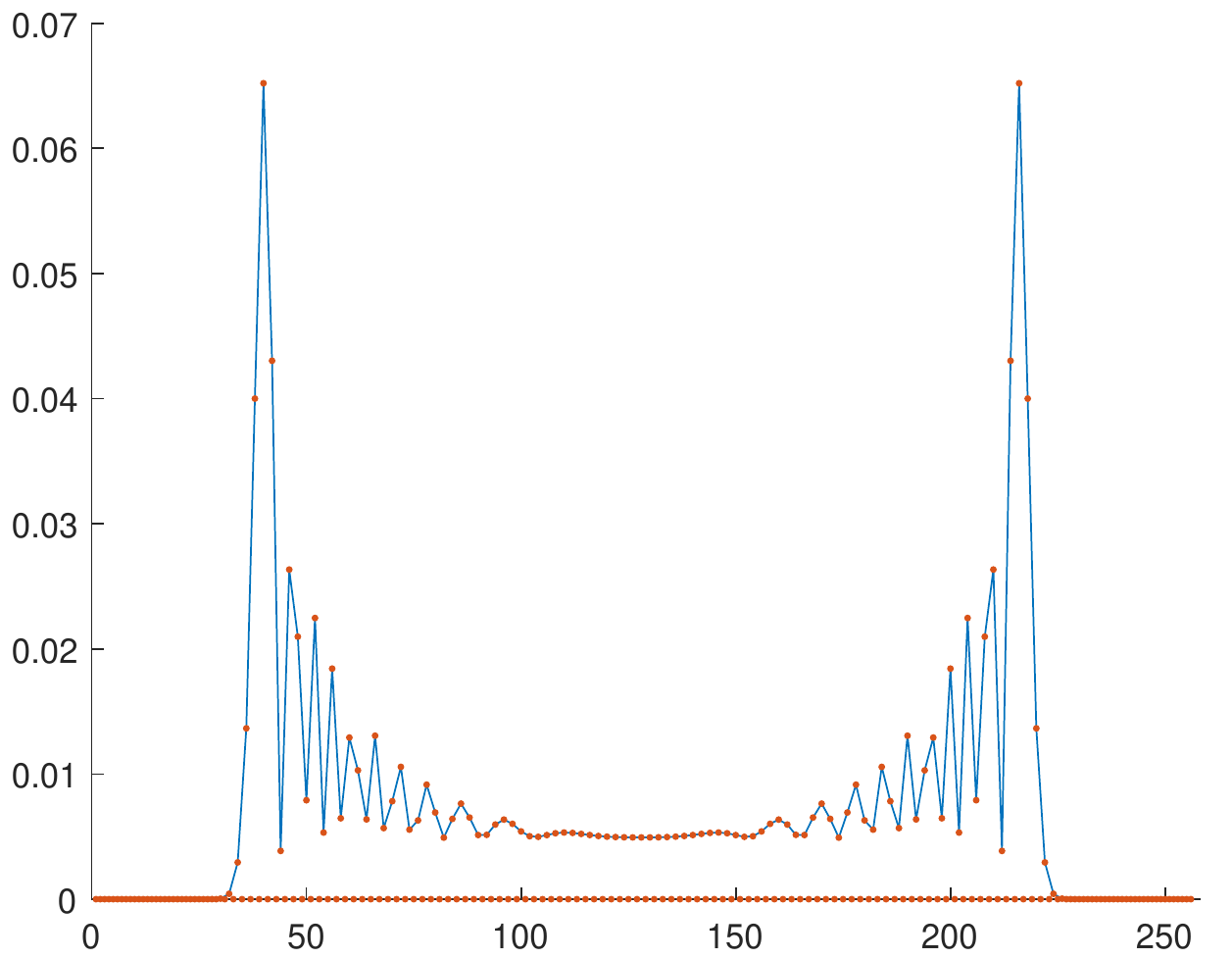}
\end{center}
\setlength{\abovecaptionskip}{-0.00cm}
\caption{\footnotesize{\textbf{The simulation of quantum random walk with the initial state is $\ket{01000000}$ (128).} The $x$ axis represent the quantum state, from $\ket{00000000}$ (0) to $\ket{11111111}$ (255). The $y$ axis represent the measurement probabilities (in the normalization units). The solid circles are the probabilities of each state. The statistical distribution can be observed by the line which linked by the nonzero circles.}} \label{fig6}
\end{figure}

We have presented a universal algorithm to preform addition and subtraction operator using a two-qubit auxiliary system above. Our algorithm has many applications and one of them is quantum random walks\cite{qrw}. Quantum random walks (QRWs) are extensions of the classical counterparts and have wide applications in quantum algorithms\cite{algorithm}, quantum simulation\cite{simulation}, quantum computation\cite{computation}, and so on\cite{others}. In standard one-dimensional (1D) discrete-time quantum walks(DTQWs), the walker's position can be denoted as $\ket{x}$ ($x$ is a integer number) and the coin can be described with the basis states $\ket{0}$ and $\ket{1}$\cite{dqrw1,dqrw2}. The evolutions of the walker and the coin are usually characterized by a time-independent unitary operator $U=TS_c{\ket{\phi}}$. In each
step, the coin is tossed by
\begin{equation}
\label{rotation}
S_c(\phi)=\left(
\begin{array}{cc}
\cos(\phi)&-\sin(\phi) \\
\sin(\phi)&\cos(\phi)
\end{array}
\right).
\end{equation}
\begin{figure}[htb]
\begin{center}
\includegraphics[width= 0.95\columnwidth]{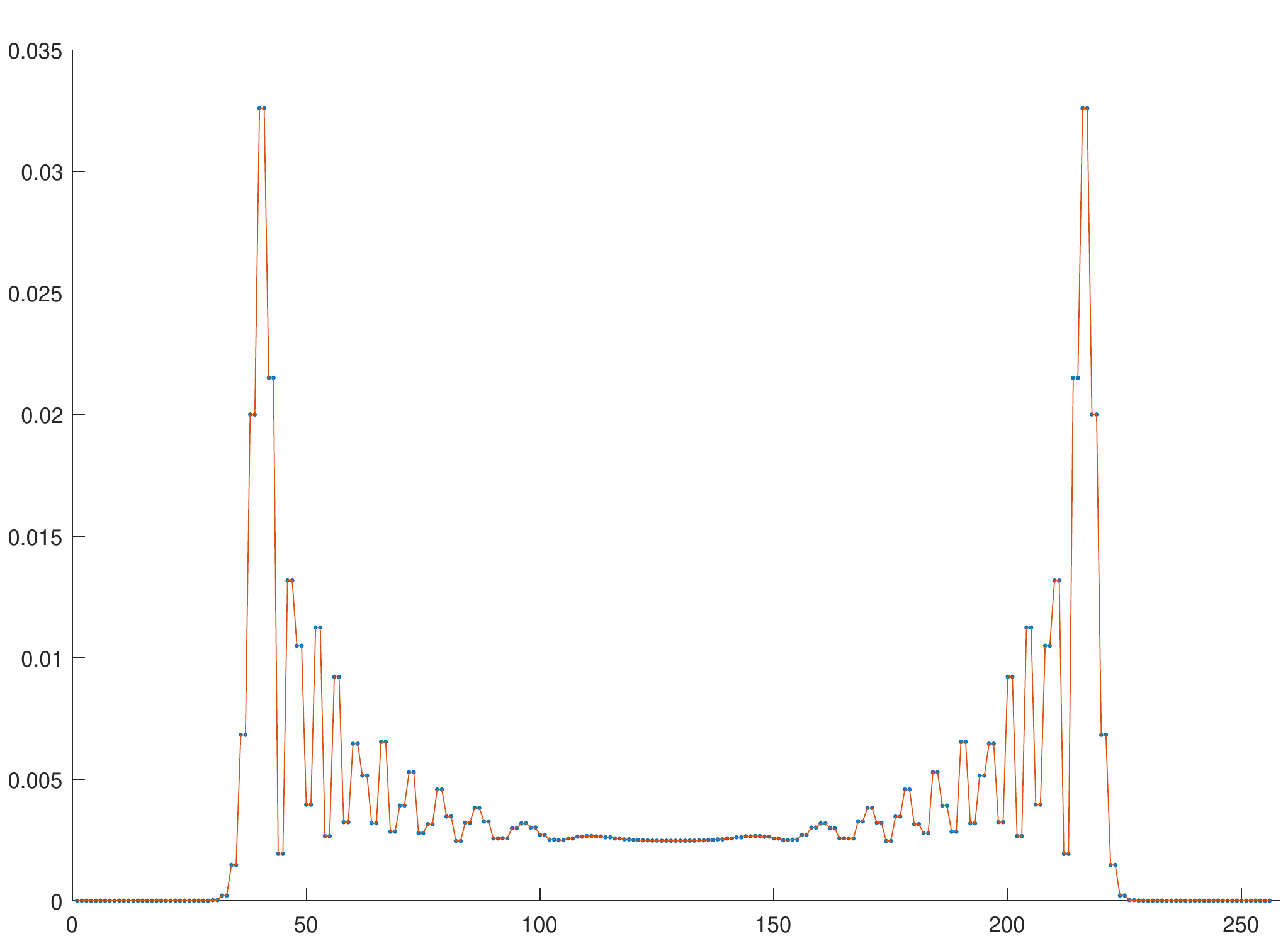}
\end{center}
\setlength{\abovecaptionskip}{-0.00cm}
\caption{\footnotesize{\textbf{The simulation of quantum random walk with the initial state is $\frac{\ket{01000000}+\ket{01000001}}{2}$(128,129).} The $x$ axis represent the quantum state, from $\ket{00000000}$ (0) to $\ket{11111111}$ (255). The $y$ axis represent the measurement probabilities (in the normalization units). The solid circles are the probabilities of each state. The statistical distribution can be observed by the line which linked by all the circles.}} \label{app2}
\end{figure}
where $\phi$ is the rotation angle and equal to $45^{\circ}$ in this work. 

The walker is shifted by $T=\Sigma\ket{x+1}\bra{x}\otimes\ket{1}\bra{1}+\Sigma\ket{x-1}\bra{x}\otimes\ket{0}\bra{0}$. In general, the result of the DTQWs with a finite number of steps is determined by the initial states of the coin and the walker as well as the operator $U$. Obviously, the operator $U$ can be realized by the algorithm we introduced above. The operator $\ket{x+1}\bra{x}\otimes\ket{1}\bra{1}$ is the addition operator when the measurement of first auxiliary qubit is $\ket{1}$. Similarly the operator $\ket{x-1}\bra{x}\otimes\ket{0}\bra{0}$ is the addition operator when the measurement of first auxiliary qubit is $\ket{0}$. So the first auxiliary qubit of our algorithm can be considered as the coin qubit of QRWs.Then we present two kinds of simulations with different initial state to demonstrate the QRWs by our algorithm. The size of the work system we choose is 8 and the auxiliary system is still a two-qubit system. The random walk step is 128 and that means we repeat the circuit in the Fig.1 for 128 times. The demonstration results with initial state $\ket{01000000}$ are shown in figure 7. We find that the probabilities of the odd state are all zero and only even state exists the probability of finding the particle. Moreover the statistical distribution has a good agreement with theory\cite{dqrw2}. As we introduced before, our algorithm can be applied on the supposition state, so we choose another initial state $\frac{\ket{01000000}+\ket{01000001}}{2}$, the simulation result is shown in Fig. 8. Obviously, now the odd state and even state are the same probabilities and the statistical distribution stay constant.

\section{conlusion}

In summary, we propose a universal algorithm to realize addition and subtraction operator for the first time. In this algorithm addition and subtraction operators can be performed by
the linear combination of unitary operations with a two-qubit
ancillary system. Moreover, the number of ancillary qubits
is independent with the size of work system. We implement
this algorithm in a four-qubit NMR quantum processor. Since
two qubits are the ancillary system, the size of work system is
four (0,1,2,3). Without loss of generality, we choose two initial quantum state that one is the superposition state, the other
is not. Our experimental results have shown good agreement
with the theoretical predictions. Moreover, the application also proves that our algorithm is a universal method to realize addition and subtraction operators with a two-qubit ancillary system. They all prove that our algorithm
can be a universal quantum control technology which can also
be utilized in other quantum physical system such as NV center, superconducting, trapped ion and so on.

This work was supported by the National Basic Research
Program of China (2015CB921002), the National Natural Science Foundation of China Grant Nos. (11175094,
91221205). Xiangyu Kong, Shijie Wei and Jingwei Wen are supported by the Fund of Key Laboratory (9140C75010215ZK65001).

\end{document}